\documentclass[%
 reprint,
 amsmath,amssymb,
 aps,
prb,
]{revtex4-2}

\usepackage{graphicx}
\usepackage{dcolumn}
\usepackage{bm}
\usepackage{placeins}
\usepackage{float}
\usepackage{svg}
\usepackage{afterpage}
\usepackage{multirow} 
\usepackage{textcomp}

\begin{document}

\preprint{APS/123-QED}

\title{Design and Operation of Wafer-Scale Packages Containing $>$500 Superconducting Qubits}

\author{Oscar~W.~Kennedy}\email{okennedy@oqc.tech}
\author{Waqas~Ahmad}
\author{Robert~Armstrong}
\author{Amir~Awawdeh}
\author{Anirban~Bose}
\author{Kevin~G.~Crawford}
\author{Sergey~Danilin}
\author{William~D.~David}
\author{Hamid~El~Maazouz}
\author{Darren~J.~Hayton}
\author{George~B.~Long}
\author{Alexey~Lyapin}
\author{Scott~A.~Manifold}
\author{Kowsar~Shahbazi}
\author{Ryan~Wesley}
\author{Evan~Wong}
\author{Connor~D.~Shelly}\email{cshelly@oqc.tech}

\affiliation{%
 Oxford Quantum Circuits, Thames Valley Science Park, Shinfield, Reading, United Kingdom, RG2 9LH}%

\begin{abstract}
Packages capable of supporting large arrays of high-coherence superconducting qubits are vital for the realisation of fault-tolerant quantum computers and the necessary high-throughput metrology required to optimise fabrication and manufacturing processes. 
We present a wafer-scale packaging architecture supporting over 500 qubits on a single 3-inch die. 
The package is engineered to suppress parasitic RF modes, and to mitigate material loss through simulation-informed design while managing differential thermal contraction to ensure robust operation at millikelvin temperatures. System-level heat-load calculations from a large wiring payload show this package may be operated in commercial dilution refrigerators. 
Measurements of the qubits loaded into the package show median $T_1$, $T_{2e} \sim 100~\mu$s ($\sim$100 qubits) alongside readout with median fidelity of 97.5\% (54 qubits) and a median qubit temperature of 36~mK (54 qubits). These results validate the performance of these packages and demonstrate that large-scale integration can be achieved without compromising device performance. 
Finally, we highlight the utility of these packages as a tool for high throughput feedback on qubit figures of merit over large sample sizes, allowing identification of performance outliers in the tails of the coherence distribution, a critical capability for informing fabrication and manufacture of high-quality quantum qubits and quantum processors.

\end{abstract}

\maketitle


\section{Introduction}
Superconducting qubits are promising candidates for fault-tolerant quantum computing~\cite{google2025quantum}. The overhead in physical qubits relative to logical qubits in error correcting codes~\cite{google2025quantum, fowler2012surface} mean that current resource estimates suggest canonical algorithms may require millions of physical qubits~\cite{gidney2025factor, yoder2025tour}. This necessitates quantum processors with larger physical-qubit counts and lower physical-qubit-gate errors. 
As packages become larger, new challenges emerge such as low frequency parasitic microwave modes hosted in the larger sample space and larger differential thermal contraction of packaging parts.
Here we consider the design and manufacture of large physical-qubit-count, low-loss, packages to support these quantum processors. The packages are designed to accept wafer-scale dies populated with superconducting qubits. We show how high-throughput qubit measurements enabled by these packages provide statistical datasets for the optimisation of qubit design and manufacture, a critical capability for continued reduction in qubit errors. 

In the first part of this manuscript we discuss the design of packages. These packages are optimised for coaxmon qubits~\cite{rahamim2017double}, which are transmons~\cite{koch2007charge} formed from coaxial capacitive pads, patterned on the opposing face of a low-loss substrate to a coaxial readout resonator with out-of-plane control and readout wiring integrated into the package. The design process considers problems common to all superconducting qubit modalities and may find relevance for other solid state qubit systems. 
We design the package to mitigate multiple packaging-induced error channels including microwave box modes~\cite{lienhard2019microwave, spring2020modeling}, packaging-material-related dielectric, conductor and seam loss~\cite{huang2023identification, huang2021microwave, ganjam2024surpassing} and Purcell decay~\cite{purcell1995spontaneous}. We optimise the construction of the package for thermal considerations, modelling the effects of differential thermal contraction of packaging parts~\cite{ekin2006experimental} and the thermal load on host cryogenic systems when using these types of package~\cite{krinner2019engineering, manifold2025thermal}.

In the second part of this manuscript we show the operation of these packages containing a 3-inch wafer scale die containing $>$500 qubit-resoantor pairs. We omit per-qubit control lines with control signals instead delivered through the multiplexed readout lines. We firstly validate the performance of this package by measuring figures of merit across statistically relevant populations of qubits. 
We present measurements of $\mathcal{O}$(100) qubits showing median-of-median coherence times $T_1, T_{\rm 2 e}\sim100~\mu$s. We measure the readout fidelity of 54 qubits with median readout errors of 2.5\% with an understanding of how to further improve these numbers. We measure the median effective qubit temperature of 36~mK over 54 qubits showing state-of-the-art, low effective temperatures for using passive-reset only. Together these metrics suggest a highly performant package which is validated by analysis of the qubits indicating that coherence limits are imposed by the material stack of the wafer-under-test. 

We consider the utility of this package as presented here, omitting per-qubit control, which can be operated as a high-throughput package requiring no wire bonding which we believe is of active interest to the community~\cite{syong2025high}. The large number of qubits measured from a single wafer gives a rich statistical dataset. Bootstrapping analysis of our large coherence dataset shows that median coherence times are determined with low error from small qubit populations, but maximum and minimum coherence times require many qubit measurements to determine with accuracy. These numbers, particularly the minimum coherence times~\cite{mohseni2024build, weeden2025statistics} expected from a process, are key system parameters for large-scale quantum computers. These packages therefore have an important role today providing feedback to iterative development cycles of processor manufacture. 

\section{Packaging Design}
\subsection{Core Architecture}

\begin{figure}
    \centering
    \includegraphics[width=\linewidth]{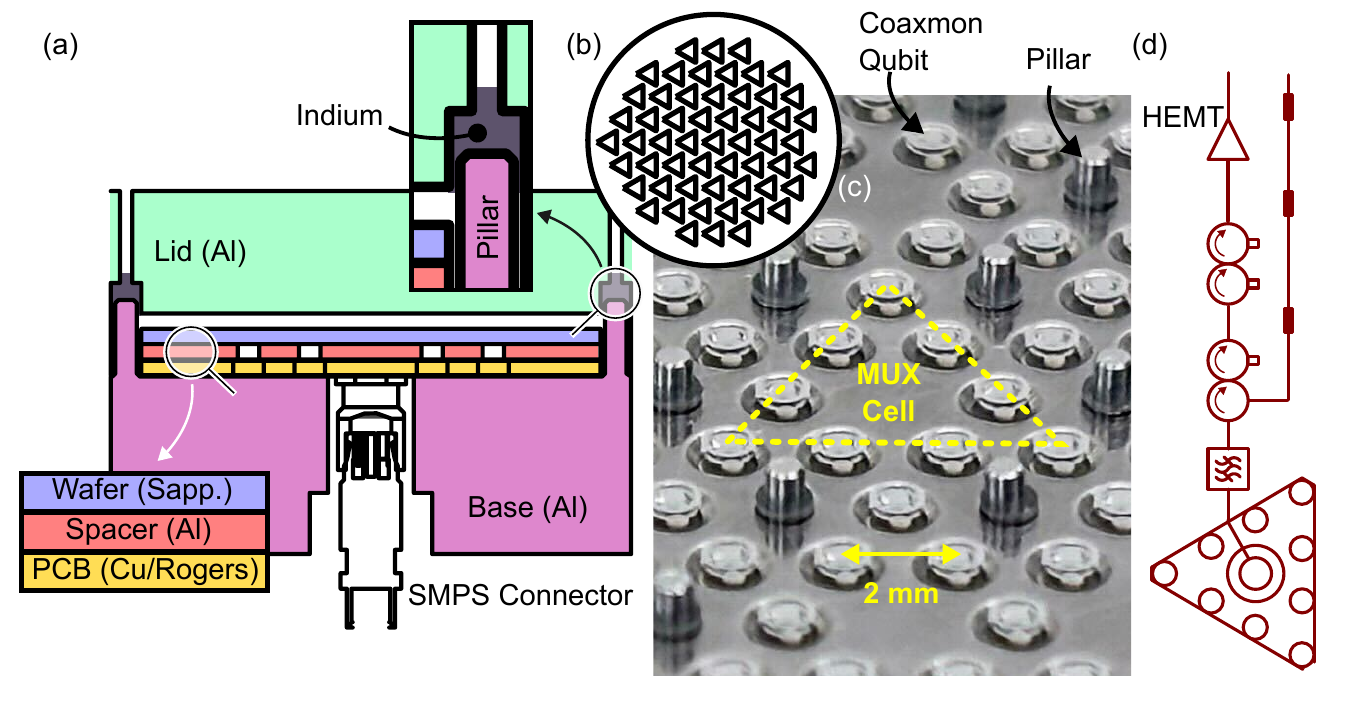}
    \caption{Schematic showing the package. (a) A cross section of the part of the package showing the superconducting lid and base, a layered stack of PCB, superconducting spacer, wafer and the joints between the lid and base where pillars push into indium. The PCB is connected out-of-plane by SMPS connectorised coaxial cables. (b) Schematic layout of 56 triangular 9-1 multiplexing cells across a wafer. (c) A photograph of coaxmon qubits and resonators sitting on top of a spacer piece. Underneath the coaxial qubits and resonators the open-circuit output ports of the PCB are visible. Pillars from the base protrude through the full PCB/spacer/wafer stack. (d) Microwave lines connected to the multiplexing cell for control and readout of qubits.  }
    \label{fig:pack}
\end{figure}

Fig.~\ref{fig:pack}~(a) shows schematics of the wafer-scale sample package which shields the qubits from environmental radiation, directs signals delivered by microwave cables to specific qubits and provides a thermal link mediated by electrons and phonons to the actively cooled stages of the dilution refrigerator~\cite{huang2021microwave}.
The cross-section in Fig.~\ref{fig:pack}~(a) shows the assembly of the package. An aluminium lid and base piece form the outer layers. Pillars are machined into the base piece, forming one solid block of aluminium, which mates with the surface piece by indium welds at the pillars.
The space between these two pieces holds a printed circuit board (PCB), an aluminium spacer piece and a wafer patterned with qubits and resonators. The pieces are bolted together ensuring electrical and thermal contact. 
At the edge of the package the lid and base are grounded through the PCB and aluminium spacer piece. A superconducting cavity is formed between the lid and the spacer piece which holds the wafer. These pieces are grounded together at the edge through an aluminium-aluminium joint and in the centre through the PCB, base plate and pillars. 
The package is mounted via an oxygen-free high-conductivity (OFHC) copper cold finger inside a light-tight OFHC copper can which has a labyrinth-style aperture to allow for atmosphere evacuation. 
This in turn sits inside a high-aspect ratio magnetic shield which suppresses stray fields by $>$20~dB. The full assembly is mounted to the base stage of a dilution refrigerator using OFHC brackets.

The PCB contains 56 triangular multiplexing cells which are shown schematically in Fig.~\ref{fig:pack}~(b). Multiplexed readout, where multiple readout resonators are coupled to a common readout line, are commonly used in superconducting circuits~\cite{chen2012multiplexed, jerger2012frequency}. 
Each multiplexed cell is a triangular filter patterned into a multi-layer PCB (copper clad Rogers dielectric) with an approximately Lorentzian line-shape and a centre frequency at $\sim$10~GHz~\cite{WA_forthcoming}. This is atypical and traditionally the multiplexing apparatus would be an on-chip superconducting element. 
Each filter has 9 pins which face superconducting resonator/qubit pairs indicated in Fig.~\ref{fig:pack}~(c) and thus has a 9-1 multiplexing ratio. The pins couple to qubits and resonators capacitively across a vacuum gap. The size of this gap is determined by the superconducting spacer piece which can be designed to set the coupling rate of the resonator to the pins (i.e. the coupling Q-factor). 
The PCB-based filters act as Purcell filters~\cite{reed2010fast, jeffrey2014fast} due to their Lorentzian lineshape which gives a passband at our readout frequencies and suppresses radiative decay over the frequency band typical for transmon qubits ($\sim$4-6~GHz). In Fig.~\ref{fig:pack}~(c) we show a photograph of the coaxmon qubits, readout resonators, spacer piece, underlying PCB and pillars which pass through these layers. We overlay a dashed line showing the footprint of one of these multiplexing cells.  

Each multiplexed readout line is connected to standard readout wiring for reflection measurements, with heavily attenuated and filtered input lines, four junctions of circulators/isolators and amplified output lines  shown in Fig.~\ref{fig:pack}~(d). 
We use hybrid filters which include a reflective filter giving a sharp roll-off and then an absorptive medium which ensures opacity up to high frequencies around the superconducting gap frequencies. 
In this setup we omit parametric amplifiers so the first stage of amplification is the HEMT amplifier.
In the configuration studied here, drive signals at qubit frequencies and readout signals at resonator frequencies are combined at room temperature before being sent to the device. The single-stage Purcell filter suppresses radiative decay but not so strongly as to completely stop strong microwave drive signals reaching the qubit to drive Rabi oscillations, meaning we can use this one port to readout and control 9 qubits. This is convenient when measuring large numbers of qubits using a small number of input lines and allows high-throughput qubit measurements. Alternatively we could modify this package to include individual per-qubit X,Y drive lines which would be integrated into the lid piece~\cite{patterson2019calibration} shown in Fig.~\ref{fig:pack}~(a) which would turn this into a QPU package, with the wiring demonstrated here acting as the readout module. 

\subsection{Box Modes}

\begin{figure}
    \centering
    \includegraphics[width=\linewidth]{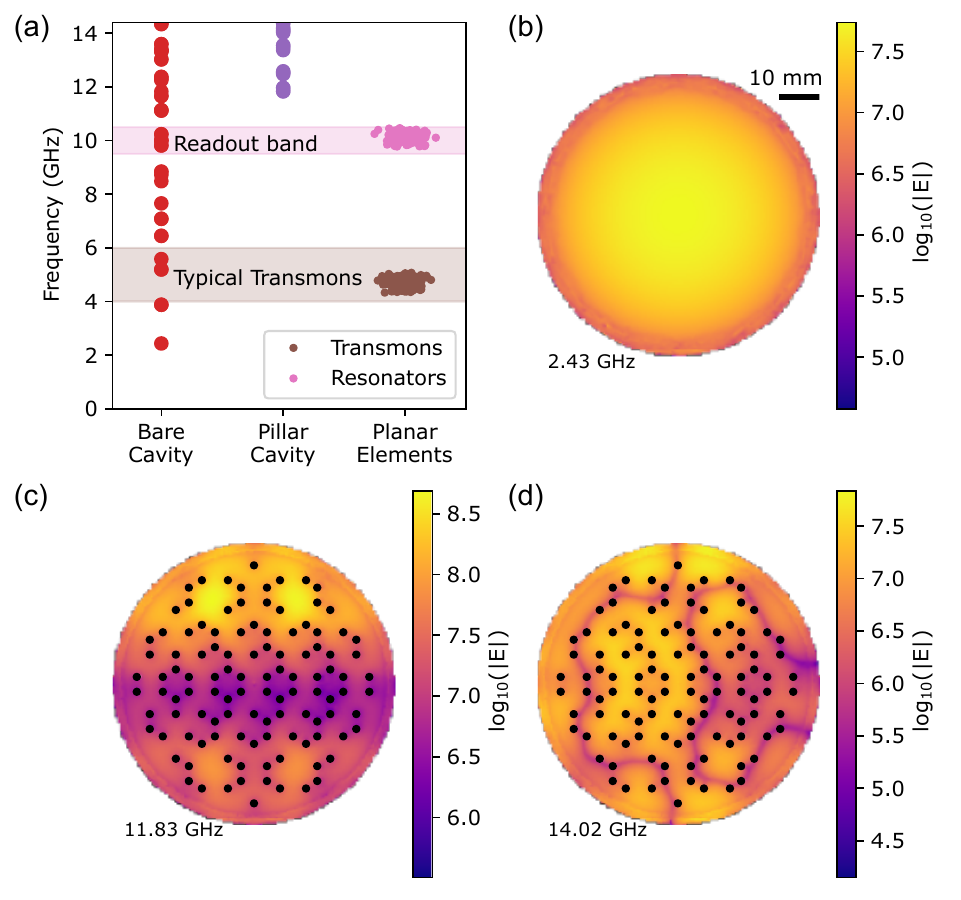}
    \caption{(a) Frequencies of RF modes of cylindrical cavities both bare and with pillars shorting the top and bottom faces. Alongside this we show the frequencies of transmon qubits measured in this work and their readout resonators which form two distinct frequency bands. (b - d) The electric field magnitude for different modes within the cylindrical cavity each holding 1~J of energy. (b) The 2.43~GHz fundamental mode of a cavity without pillars. (c) The 11.83~GHz fundamental mode of a cavity with pillars present as indicated by black dots. Here the electric field maxima occur towards the edge of the wafer. (d) A 14.02~GHz mode with substantial electric field strength within the forest of pillars showing a mode that extends into the pillars.  }
    \label{fig:modes}
\end{figure}

The monolithic superconducting die sits inside a low aspect ratio, approximately cylindrical, superconducting cavity. The package enclosure is made from machined pieces of aluminium which form the cavity between the lid and the spacer piece. 
The diameter of the cavity required to host a 3" wafer results in a family of box modes indicated by red points in Fig.~\ref{fig:modes}~(a) with a fundamental mode at $\sim$2.4~GHz with the E-field distribution of the mode shown in Fig.~\ref{fig:modes}~(b).
The family of modes span the frequency band of typical transmons and readout resonators as well as specifically those used in this work. 
These box modes could allow for Purcell enhanced radiative decay of the qubits into resonant cavity modes~\cite{purcell1995spontaneous, huang2021microwave}, control and readout cross talk mediated by box modes~\cite{spring2020modeling} and measurement-induced dephasing by these box modes~\cite{sheldon2017characterization}.

To mitigate these deleterious effects we introduce an array of pillars which provide a galvanic short between the base and lid of the cavity~\cite{spring2020modeling}. These pillars pass through apertures machined into the monolithic die as in Ref.~\cite{spring2022high, acharya2025integration}. The array of pillars acts as a microwave metamaterial and push the fundamental box mode up to $\sim$12~GHz which has E-field maxima at the edge of the cavity as shown in Fig.~\ref{fig:modes}~(c). Modes at $\sim$14~GHz and above have large amplitude across the centre of the cavity as shown in Fig.~\ref{fig:modes}~(d). The family of modes hosted in this cavity shown in Fig.~\ref{fig:modes}~(a) no longer overlap the qubit or readout frequency bands, reducing the magnitude of the negative effects they might otherwise cause. We show the spatial distribution of the E-field for first 16 modes of the cavity with and without the shunting pillars in the supplementary materials.    

\subsection{Packaging Loss Budget}

\begin{figure*}
    \centering
    \includegraphics[width=\linewidth]{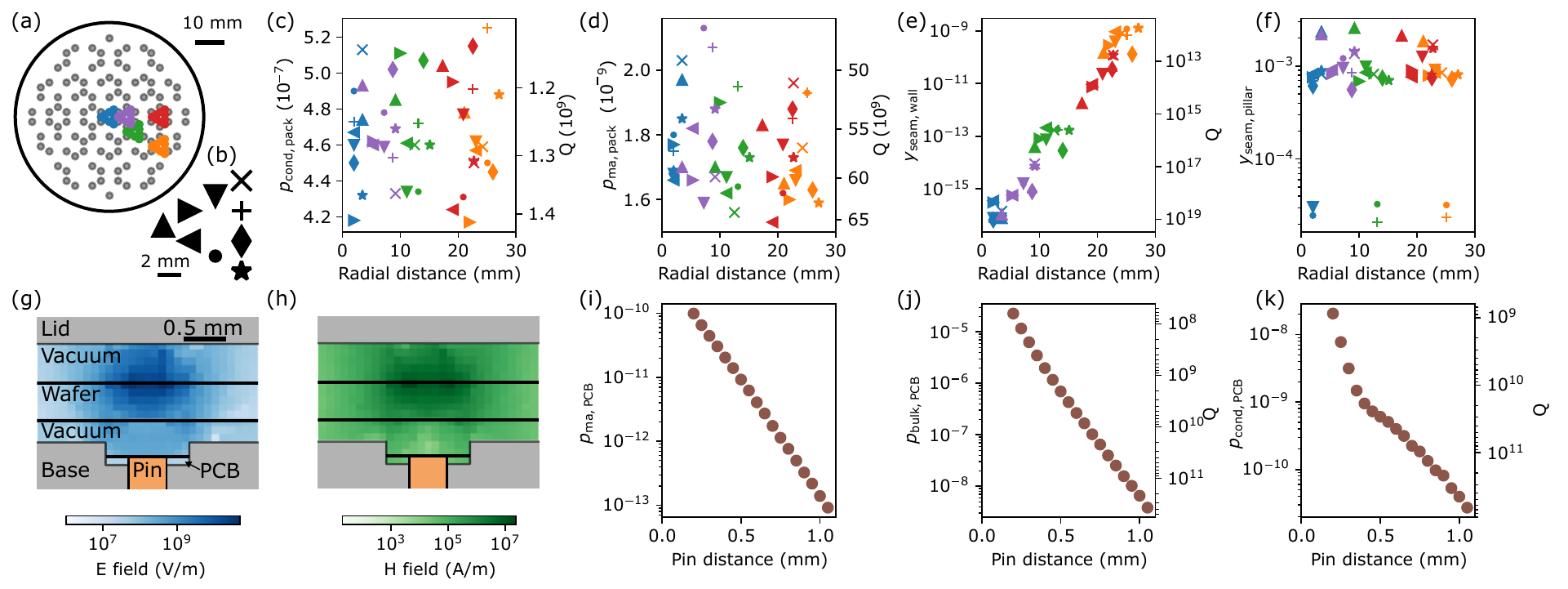}
    \caption{Finite element simulations of participation ratios for different loss channels arising due to packaging materials. 
    (a) Map of the package showing the qubit locations used in finite element simulations shown in panels (C-f). Grey circles indicate locations of pillars and coloured markers indicate qubits on different multiplexing cells. (b) A key indicating the markers used for individual qubits on a multiplexing cell. The results of loss simulations are shown in (c) conductor loss of the package walls (d) metal air dielectric loss of the package walls (e) seam loss from the side walls where package pieces join (f) seam loss from the pillar terminating in the package lid. 
    Cross-section of the package showing (g) E-field and (h) H-field for an example simulation with the distance between the readout pin and the bottom of the monolithic wafer indicated. The E-field strengths are shown when 1~J of energy is stored in the mode. 
    We simulate different participation ratios which vary with this distance including (i) metal air dielectric loss of the PCB metal surface (j) dielectric loss in the PCB bulk dielectric (k) the conductor loss from the exposed surface of the PCB.
    (c-d, j-k) Have a second y-axis indicating the Q factor limit from these loss channels which are determined using material loss-factors from literature values~\cite{lei2023characterization, calatroni2019cryogenic, mazierska2005loss}. (f, i) Do not include a second y-axis due to the lack of literature values governing these materials at relevant operating conditions.}
    \label{fig:losses}
\end{figure*}

Microwave loss in the materials which make up the packaging for superconducting qubits can suppress their coherence times. For instance, using normal conducting packaging components has been shown to reduce quality factors of microwave resonators~\cite{huang2023identification}. 
Here we consider packaging materials related loss channels of (i) dielectrics in the packaging materials, (ii) ohmic losses in metal components,  (iii) finite surface conductivity in superconducting components and (iv) seam losses between packaging components. These loss rates can be simulated by finding their respective energy participation ratio (EPR)~\cite{gao2008experimental, wenner2011surface, wang2015surface, ganjam2024surpassing}. The EPR is found by integrating the energy stored in the electric or magnetic field of the qubit mode within a specific region of material and normalising it to the total energy stored in the mode.

To find the dielectric loss rate we compute 
\begin{equation}
    p_E = \frac{\int_{\rm Dielectric} \epsilon(\boldsymbol{r})|\overrightarrow{E}(\boldsymbol{r})|^2 {\rm d}V}{\int \epsilon(\boldsymbol{r})|\overrightarrow{E}(\boldsymbol{r})|^2 {\rm d}V}
    \label{eq:E_EPR}
\end{equation}
where $\epsilon(\boldsymbol{r})$ is the spatially varying dielectric constant of the constituent materials. For surface dielectrics the volume integral in the numerator is typically replaced by a surface integral which is then multiplied by the dielectric thickness. If the surface layer is a thin dielectric coating on the surface of a metal (i.e. a surface oxide), the integration surface is defined to be a surface in the vacuum just outside the metal. In this case, the continuity of the displacement field means that the calculated $p_E$ in Eq.~\ref{eq:E_EPR} is then divided by $\epsilon_{r, \rm Ox}^2$. 

Conductor loss occurs due to screening currents induced to flow in the metallic walls by the magnetic component of the mode. The participatio ratio of this loss mode is computed by 
\begin{equation}
    p_{\rm cond} =  \frac{\lambda \int_{\rm surf} |\overrightarrow{H_{||}}|^2 {\rm d}S}{\int |\overrightarrow{H}|^2 {\rm d}V}
    \label{eq:conductor}
\end{equation}
where $\overrightarrow{H_{||}}$ is the component of magnetic field parallel to the surface of the package which induces the screening currents. The Q-factor of a mode experiencing only conductor loss is given by $Q_{\rm cond} = R_s/\omega\mu_0\lambda p_{\rm cond}$ where $R_s$ is a surface resistance, $\mu_0$ is the permeability of free space and $\lambda$ is the depth of field penetration into the metal.

Seam loss occurs when screening currents in the package flow across a seam between two conducting parts and finite resistance present at the seam causes Ohmic losses. Seam admittance is computed as
\begin{equation}
    y_{\rm seam} = \frac{\int_{\rm seam} |\overrightarrow{J}\times \hat{l}|^2 {\rm d}l}{\omega\int \mu_0|\overrightarrow{H}(\boldsymbol{r})|^2 {\rm d}V}
    \label{eq:seam}
\end{equation}
where the Q-factor of a cavity experiencing only seam loss is given by the ratio $Q_{\rm seam} = g_{\rm seam}/y_{\rm seam}$. The term $g_{\rm seam}$ is the seam conductance, a property which depends on the materials used, how they are processed and assembled. The integral in the numerator is more readily computed in finite element (FE) software as $\int_{\rm seam} = |\overrightarrow{H_{||}}|^2dl$ where  $\overrightarrow{H_{||}}$ is the component of the magnetic field parallel to the seam~\cite{krayzman2022thin} and the integral in the denominator runs over all space.

Fig.~\ref{fig:losses} shows the losses from the packaging for qubits at different locations within the package. 
Fig.~\ref{fig:losses}~(a) shows the locations of qubits considered in these simulations - qubits with different multiplexing cells are denoted by their different colours. Fig.~\ref{fig:losses}~(b) shows a key identifying the markers used for different qubits on each multiplexing cell.
Each of these qubits has the same design frequency of 4.5~GHz but has different proximity to neighbouring pillars and the packaging side walls meaning their loss varies.  
In Fig.~\ref{fig:losses}~(c - f) we show respectively (c) loss from screening currents in the packaging walls (d) dielectric loss from the package walls (e) seam loss from the Al/Al joints at the edge of the package and (f) seam loss from the Al/In joints where pillars mate with the top of the package enclosure. We use Al-5083, a machineable grade of aluminium and tabulate key material parameters used in these simulations in the supplementary materials referencing~\cite{faber1955penetration, lei2023characterization, wenner2011surface, de1989oxidation,calatroni2019cryogenic, petousis2017high, mazierska2005loss, ratzinger2022anomalous, RogersData}.

The loss in qubits due to packaging screening currents and dielectric loss has little spatial dependence as shown in Fig.~\ref{fig:losses}~(c,d). The seam loss from the sidewalls of the package, shown in Fig.~\ref{fig:losses}~(e), increases strongly for qubits closer to the walls, but remains small giving Q-factor limits of $\sim10^{12}$ greatly in excess of record material-limited qubit quality factors $\sim10^7$~\cite{bland2025millisecond}. This is because the majority of the screening currents flow through pillars which are much closer to the qubits than the sidewalls as shown by the larger values in Fig.~\ref{fig:losses}~(f). The seam loss from pillars is similar for most qubits with some low-lying values of $y_{\rm seam, pillar}$. The qubits with lower loss are further from the nearest pillar, $\sim$3.4~mm versus $\sim$2~mm for qubits with higher loss.

As shown in Fig.~\ref{fig:pack}~(a) a piece of indium is placed in a recessed hole at this joint to provide a superconducting joint. Seams from In/In joints have been reported with $g_{\rm seam, In/In} \ge 2\times10^{10}~\Omega^{-1}{\rm m}^{-1}$~\cite{lei2020high} but In/Al joints have been reported to form non-superconducting intermetallics~\cite{paradkar2025superconducting}. Combining loss simulations and coherence measurements presented later in this work we can provide a lower bound on the seam conductivity of this interface $g_{\rm seam, Al/In} > 3\times10^3~\Omega^{-1}{\rm m}^{-1}$ assuming that qubit loss is entirely dominated by this seam. This is $\sim 4 \times$ larger than the seam conductivity from aluminium/aluminium joints~\cite{lei2023characterization, brecht2015demonstration} indicating that this is a low loss joint. We note that the true conductivity of this seam is likely higher than this lower bound given that, as shown later in the manuscript, we are able to attribute most qubit loss to on-chip loss tangents. Either the bulk, machineable grade of aluminium does not form the intermetallic with the indium upon cold-welding, or the increased contact area formed with indium more-than compensates for any increase in resistivity. 

In our package we can adjust the distance between the surface of the PCB and the base of the wafer by choosing different spacer pieces. These changes cause the qubit to move relative to the lossy PCB materials including silver-coated copper traces and Rogers RT Duroid 5880 dielectric. In Fig.~\ref{fig:losses}~(g,h) we show cross-sections of the E- and H-field of the qubit modes and their distribution within the package and label the different regions. 
In the simulations presented Fig.~\ref{fig:losses}~(i-k) we show the dielectric loss from the oxide surface of the metal traces, dielectric loss from the bulk PCB material and loss from screening currents in the PCB traces. We compute this as a function of the distance between the bottom of the wafer and the top of the PCB and which we refer to as the pin distance. 
We tabulate material properties used in these simulations in the supplementary materials drawing on values from~\cite{de1989oxidation, calatroni2019cryogenic, mazierska2005loss}.

Loss increases as pin distance decreases and the qubits move closer to the lossy PCB materials. The external quality factors of the readout resonators will also drop as the pin distance decreases allowing for faster readout of the qubit state meaning there is a trade-off between the coherence achievable and the speed of readout. Using realistic material properties we find that package-losses remain low even as the pin approaches the qubit - a promising prospect for improved fast-readout with more strongly-coupled readout resonators. 

We compute the total packaging-induced loss by summing the different loss channels presented in Fig.~\ref{fig:losses}. We omit loss from the surface oxide of PCB metal traces and seam loss from In/Al joint at the top of the pillars due to a lack of available loss-rates describing these materials in the literature. 
For pin distances of 0.5~mm as used in the following experiments we compute the minimum qubit Q-factor to be $>$500~million ($T_1 = 20$~ms for a 4.5~GHz qubit) and for the smallest pin distances simulated, 0.2~mm, we find Q-factors $\sim$35~million ($T_1 = 1.2$~ms for a 4.5~GHz qubit). Relative to even state-of-the-art qubit coherence times, these Q-factors are large which is promising for a performant quantum computer and for a high-throughput testing where materials-related figures of merit will not be affected by the packaging. 

\subsection{Thermal Contraction}

\begin{figure}
    \centering
    \includegraphics[width=\linewidth]{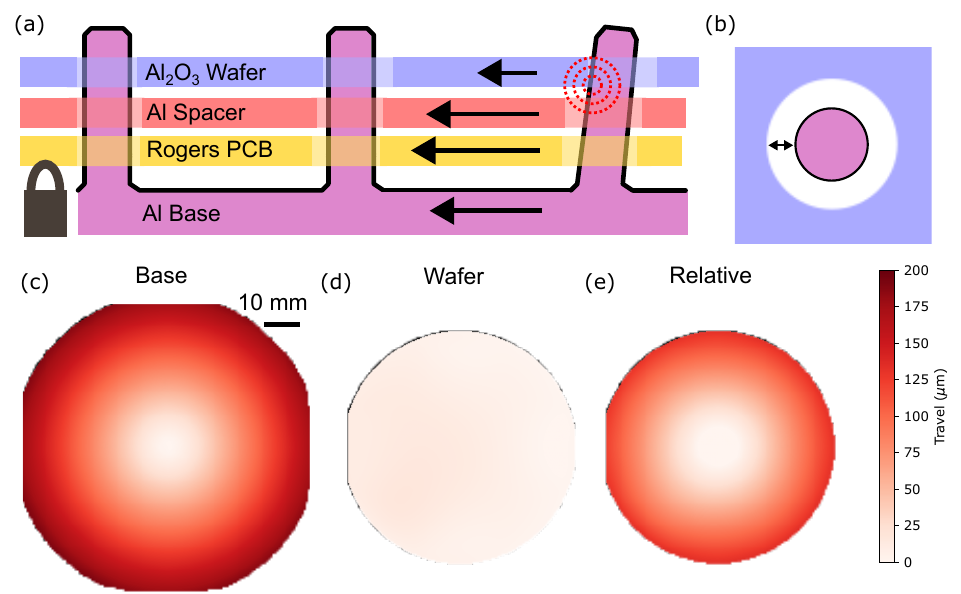}
    \caption{(a) Schematic showing the effects of differential thermal contraction upon cooling, where parts may collide with one another causing fine features to fail. The system is anchored on the left hand side  (b) A birds-eye view of a pillar protruding through an apperture in a wafer showing the tolerancing of the aperture around a shunting pillar. We show the thermal contraction of (c) the Al base plate (d) the wafer (e) differential between (c) and (d) showing the relative travel of the pieces.}
    \label{fig:thermals}
\end{figure}

Qubits are measured at cryogenic temperatures, but packages are assembled at room temperature. The components that make up the package will differentially contract upon this $\sim$300~K of cooling which means that we must design the package both for room temperature assembly and cryogenic operation. At cryogenic temperatures we require that (i) the package remains intact and (ii) PCB outputs remain aligned to qubit/resonator pairs. An example situation where (i) fails would be that the monolithic die contracts differentially to the package, it collides with metal pillars, breaking them, compromising the box-mode mitigation strategy shown schematically in Fig.~\ref{fig:thermals}~(a). If (ii) fails then we may see modified coupling rates to qubits and resonators, and in extreme cases be unable to couple to the quantum elements. We can design the package with knowledge of how elements will contract, for instance, in Fig.~\ref{fig:thermals}~(b) where we show tolerancing of wafer apertures around metallic pillars allowing for relative travel. 

Example results of finite element modeling of cryogenic contraction are shown in Fig.~\ref{fig:thermals}. In the simulation, a monolithic 3\textquotesingle\textquotesingle \space sapphire wafer with holes drilled through the substrate is placed over an aluminium base plate with pillars extending vertically through the holes in the wafer. 
Mechanical and thermal strain are then computed by assigning temperature dependent coefficients of thermal expansion and Young's modulus to the different materials. The simulations are stabilized using a weak-spring technique with friction governing motion between parts. The simulation computes the effect of cooling the system to 4~K with results shown in Fig.~\ref{fig:thermals}~(c-e). At the edges of the wafer there is a relative travel of $\sim$120~$\mu$m. By ensuring that gaps indicated in Fig.~\ref{fig:thermals}~(b) are larger than this we design against pillar failures. Selecting packaging materials with similar coefficients of thermal expansion to the wafers used to form the monolithic die will allow smaller tolerances and larger monolithic dies. 

\subsection{Thermal Load Simulations}

In this work we operated the package measuring a subset of the total available qubits. Here we perform thermal modeling considering use-cases where the package is fully wired to measure all qubits. We consider additional functionality including per-qubit control lines and superconducting parametric amplifiers. These simulations give confidence that the fully-wired package can be loaded into a commercial dilution refrigerator and reach the the necessary operating temperatures of $<$~20 mK. 

The operating temperature of dilution refrigerators is constrained by heat load contributions from active components, in this case HEMTs, passive heating due to thermal conductivity of wires and dissipative heating from the microwave control and readout tones~\cite{krinner2019engineering}. 
Following the methodology in Ref.~\cite{manifold2025thermal} we calculate the heat load on the system for a fully-connected package including per-qubit drive lines which requires a wiring payload which delivers a passive heat load to the system. The payload consists of 504$\times$60~dB attenuated drive lines, 56$\times$ readout in/out configurations shown in Fig.~\ref{fig:pack}~(d) and 56$\times$50~dB attenuated lines for parametric amplifier pump lines. 
We then consider the dissipative load to the system which is determined by the drive signals applied to the package in different likely operating regimes - i.e. we must drive harder in a high-throughput operation to account for signal attenuation through the Purcell filter. In our modelling these considerations translate to different conservative estimates on effective continuous drive power per qubit of -64~dBm in a high-throughput configuration omitting drive lines and -78~dBm when they are included. 
We additionally estimate a typical CW pump power for each parametric amplifier to be $\sim$-60~dBm~\cite{macklin2015near}. 

The results from simulations of both operating regimes are reported in Table S2 of the supplementary materials. Both modes of operation are found to be compatible with the modeled dilution refrigerator. Stage temperatures and helium circuit pressures are within system limits and able to provide a suitable environment for qubit measurement. The highest estimated MXC heat load predicted is $\sim$3~$\mu$W, small relative to the typical cooling power of 25-30~$\mu$W at 20 mK for commercially available dilution refrigerators. This overhead can account for non-idealities in the real payload vs the modeled payload and allow an increase in measurement cadence if enabled, for example, by active reset protocols.

\section{Characterising the Monolithic Die}
\subsection{Measuring $\mathcal{O}$(100) qubits}

Operating our package omitting per-qubit control lines we measure 105/108 qubits available on 12 multiplexing cells. We use QBlox control hardware with qubit control performed using 3 QCM-RF II cards and readout using 6 QRM-RF cards allowing us to address 6 lines simultaneously. We couple the output of the control and readout cards at room temperature using a directional coupler sending both control and readout signals to the multiplexing elements in the PCB. 

A typical workflow to characterise single qubit properties is spectroscopically identifying the frequency of a readout resonator, spectroscopically identifying the corresponding qubit frequency, iteratively performing Rabi and Ramsey sequences to calibrate the power and frequency of pulses for single qubit gates, optimising readout parameters, measurements of coherence times ($T_1$, $T_{2e}$, $T_{2}*$) with optional extensions of fine-tuning single qubit gate pulses, measurements of higher excited states and more. In order to do this for large numbers of qubits in a moderate amount of time, these routines must be automated and performed simultaneously upon many qubits. 
Automatic calibration routines are available as commercial and open source~\cite{egger2014adaptive, pasquale2023towards} software products. Here we use a simple in-house automation routine which performs sequential calibrations optimising figures of merit and flagging any failed calibrations. Calibrating operations on qubits sequentially means the length of the calibration protocol scales linearly with the number of qubits being calibrated. Batching qubits for simultaneous calibration allows a calibration protocol that scales with the number of batches rather than the number of qubits. The 9-1 multiplexing architecture naturally lends itself towards using 9 batches of qubits, where one qubit on each multiplexed line is calibrated at a time. In this work we use two batches of six qubits, where this is limited by the size of the microwave control hardware which could be straightforwardly increased.

\subsection{Qubit Readout}
\begin{figure}
    \centering
    \includegraphics[width=\linewidth]{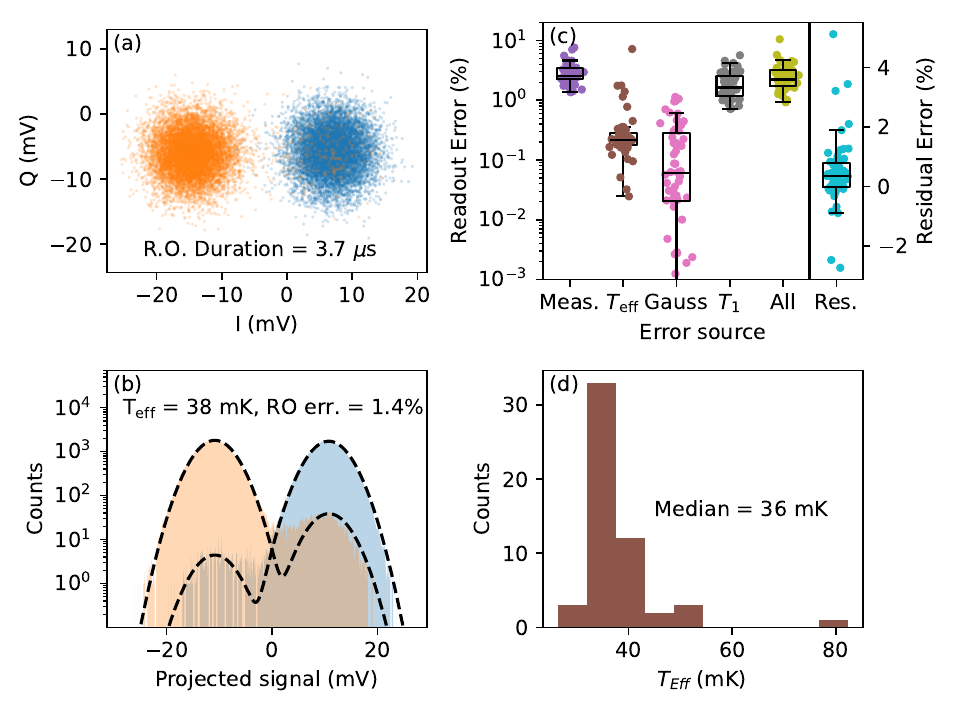}
    \caption{Readout of 54 qubits. (a) Results from reading out one of the qubits showing 10k of the total 100k measured shots in the IQ plane. (b) shows the full data set projected onto the line of maximum state-discrimination. (c) A scatter plot showing the readout error from 54 qubits with optimised readout. We truncate the y axis for improved visibility but have overlap errors as small as $5\times10^{-7}\%$. A second linear y-axis is shown with the residual error (measurement - predicted). (d) A histogram of the effective temperatures of the same qubits extracted from the ground state histogram. Equivalent plots to (a) and (b) are shown for all qubits in the supplementary materials demonstrating that the results presented here are representative. }
    \label{fig:readout}
\end{figure}
We optimised the readout fidelity for a subset of 54 qubits by ensuring that the passive reset time greatly exceeded the $T_1$ time of the qubits ($\sim8\times$) before iteratively optimising the readout pulse frequency, duration and amplitude.  We show an example of the raw IQ data after readout optimisation in Fig.~\ref{fig:readout}~(a). To collect these data we prepared the ground state by waiting for passive reset, then played our readout pulse and repeated this full motif 100k times. Next we prepared the ground state as above, excited the qubit using two $\pi_x/2$ pulses and played our readout pulse, again repeating this 100k times. 
The same data are projected onto the line of maximum discrimination (i.e. the line between the centre of the two clouds in IQ space) and translated so that the point half-way between the two centroids is at 0~mV which is shown as a histogram in Fig.~\ref{fig:readout}~(b). 

We simultaneously fit a double Gaussian to the two data sets where the seven free fit parameters are the common standard deviation, two centres and four amplitudes. We compute the readout error by finding the ratio of the number of points which fall on either side of 0~mV, our state discriminator, for both the ground and excited state and present an averaged value. We compute the effective qubit temperature considering only the ground state and interpret the relative amplitude of the two Gaussian peaks as a Boltzmann distribution. 

We show the distribution of the readout error and temperature for the set of 54 qubits in Fig.~\ref{fig:readout}~(c) and Fig.~\ref{fig:readout}~(d) respectively. 
Fig.~\ref{fig:readout}~(c) shows the results of error budgeting relative to the measured readout error. 
The error induced by finite effective qubit temperature is determined by the relative amplitudes of Gaussian peaks when preparing the qubit in the ground state shown in Fig.~\ref{fig:readout}~(b). This accounts for a median 0.3\% error. 
The error caused by overlapping clouds in the IQ plane (labelled Gauss) accounts for a median 0.1\% error which aligns with the observation that the clouds of points look well separated in Fig.~\ref{fig:readout}~(a) and for all qubits with the full data shown in the supplementary information.
Following~\cite{bengtsson2024model} we compute the errors arising from qubit decay during readout as $T_{\rm m}/2T_1$ where $T_{\rm m}$ is the duration of the measurement. This only can occur when the qubit is in the excited state, so we divide this number by two as here we show avereraged errors for preparing the qubit in the ground and excited states. We find that this is the leading term in readout errors, accounting for $\sim$1.8\% error. These can be mitigated by faster readout, for instance using parametric amplifiers and by improving coherence times of qubits. 
Fig.~\ref{fig:readout}~(c) shows the sum of all readout error terms and the residuals (measurement - prediction) and find that the estimated median value is similar to that which we measure. The residual errors have median residual error of 0.3\% indicating that the error budgeting predicts the dominant terms. This residual error may arise due to readout error mechanisms not described by this relatively simple error budgeting, such as measurement induced state transitions~\cite{reed2010high}. Additionally some errors to the budgeting may arise as we use median $T_1$ times for the qubits measured at different times to when the readout optimisation is performed and qubit coherence times may fluctuate~\cite{burnett2019decoherence}  which likely is the cause of the negative values in the residual errors. 

Some straightforward steps would allow us to reduce the leading term in our errors, $T_1$ decay during the relatively long readout pulses (median 6.2~$\mu$s). 
Firstly, the distance between the readout PCB and the qubit chip can be reduced, resulting in stronger coupling between the Purcell filter and the readout resonators. In this work our $\sim~10$~GHz resonators have $Q_{\rm ext} \sim 30$k ($\kappa_{ext}/2\pi \sim$~0.3~MHz) whereas values up to $\kappa/2\pi~\sim$~40~MHz have been used in examples of fast high fidelity readout~\cite{spring2025fast}.
Secondly, superconducting parametric amplifiers can be included in the readout chain amplifying signal with lower added noise. This has been widely used to increase measurement efficiency in superconducting circuits~\cite{murch2013observing}.
Both of these techniques will be crucial when operating this as a readout module for a fault tolerant quantum computer.
Using the current configuration, our readout error could be further improved with readout shelving techniques~\cite{chen2023transmon} or driven reset~\cite{geerlings2013demonstrating}.

The effective temperature of these qubits matches best-in-class for passive reset~\cite{jin2015thermal} and is measured over an ensemble of 54 qubits in our large package. The temperature of our qubits could likely be further reduced using active/driven reset protocols~\cite{riste2012initialization, geerlings2013demonstrating} without modifying the qubit circuitry. 
We note that qubit temperature measurements were performed after the system had been at millikelvin temperatures for approximately 6 months. Some long-tailed transients, such as the rate of strain-relief-induced phonon bursts~\cite{yelton2025correlated} have been shown to drop over time which may contribute to the low temperature measured. 

\subsection{Coherence Statistics}
\begin{figure}
    \centering
    \includegraphics[width=\linewidth]{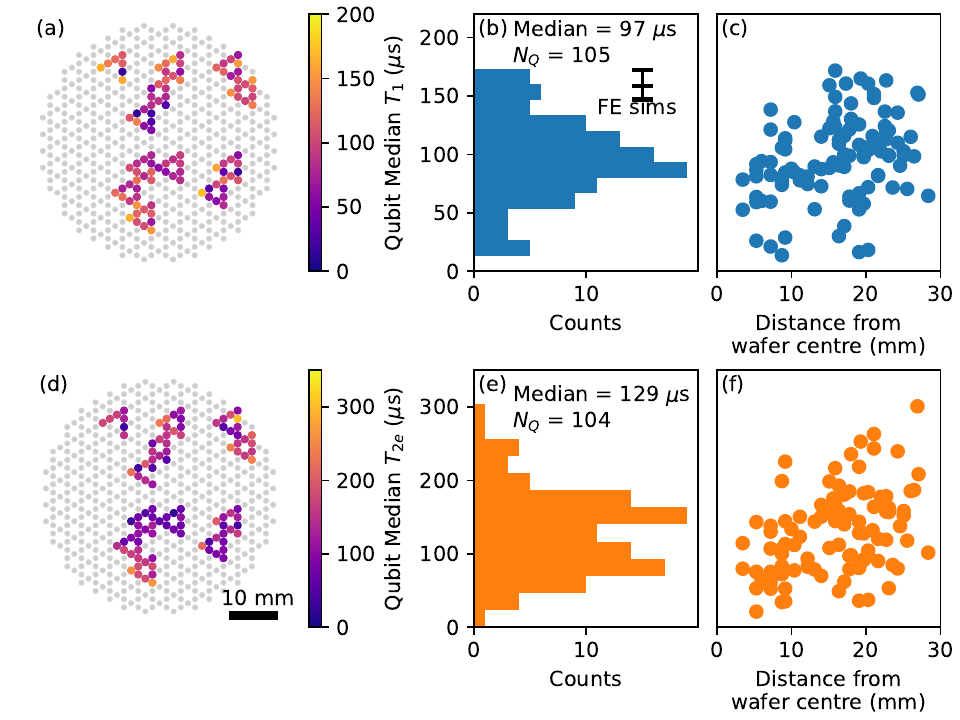}
    \caption{Qubit coherence from the monolithic die showing $T_1$ (a - c) and $T_{2e}$ (d - f). (a, d) We show spatially resolved data where grey points indicate a qubit on the monolithic die for which we do not present a coherence time and the colured dots show the relevant coherence times of the qubit according to the colour associated colour bar. (b, e) Histograms of coherence times of all measured qubits. Each value in both maps and histograms is a median value for the qubit. The median of medians and number of qubits measured are indicated in the plot. In panel (b) we indicate the predicted coherence times using finite element simulations and the loss tangents from the aluminium surfaces and the bulk dielectric drawn from literature. (c, f) The coherence times plot as a function of distance from the centre of the wafer as an alternate visualisation for spatial dependence. }
    \label{fig:coherence}
\end{figure}

We measure the coherence times $T_1$ and $T_{2e}$ for qubits. For each qubit we measure each value 50 times. Each curve is fit to an exponential with a single fit parameter, the coherence time in question. We compute $R^2$ to characterise the goodness of each fit and filter these so that $R^2 > 0.75$ before accepting the fit. We then compute the median of fits that pass the $R^2$ filter and report that as the qubit median. This ensures that we have time-averaged data for each qubit sampling temporal fluctuations of qubit parameters~\cite{burnett2019decoherence}.

In Fig.~\ref{fig:coherence} we show these data with both spatial maps (a, d) of qubit coherence times as well as histograms (b, e). We show a median $T_1 = 97~\mu$s and $T_{2e} = 129~\mu$s. The spatial maps in Fig.~\ref{fig:coherence} (a, d) do not clearly show regions of modified coherence times. In Ref. ~\citenum{van2024advanced} they consider the radial dependence of coherence times in order to assess their manufacture process. Here we have access to the same data shown in Fig.~\ref{fig:coherence}~(c, f) from our monolithic die without having to prepare and measure multiple dies and see a trend of slightly increasing coherence at the edges of the wafer. This may arise due to processing inhomgeneity or be spurious, something that may be determined by testing of further dies. In the supplementary materials we consider the Pearson correlation coefficient to quantify spatial correlations and conclude, that in agreement with a casual inspection of Fig.~\ref{fig:coherence} (a, d) there are no significant spatial trends. 

Using finite element simulations of the qubit geometry we are able to infer $T_1$ times limited by the loss tangents of the aluminium pads and bulk substrate which are shown in Fig.~\ref{fig:coherence}~(b). Using best-in-class loss tangents from aluminium on sapphire~\cite{ganjam2024surpassing, read2023precision} we find $T_1$ times similar to the best qubits measured here. This aligns with the simulations presented in Fig.~\ref{fig:losses} which suggest that the packaging induces small amounts of extra loss to the qubits and that the qubit materials limit these figures of merit. It is likely that the lower-coherence times in other qubits are caused by a combination of loss from strongly-coupled two-level-systems~\cite{simmonds2004decoherence}, material inhomegeneity or lossy residues on some regions of the device~\cite{mahuli2025improving}.

As coherence times of individual qubits fluctuate over time~\cite{burnett2019decoherence}, different metrics representing coherence times can be quoted, the highest, lowest and median value as well as a, potentially complicated, distribution. 
The distribution in coherence times of single qubits is interpreted as being caused by ubiquitous two level systems, arising in the qubit materials and coupling to the qubit. These have been successfully characterised by adding a tuning parameter, either to the qubit~\cite{simmonds2004decoherence, klimov2018fluctuations}, the defect~\cite{grabovskij2012strain, dane2025performance} or both~\cite{chen2025scalable, bilmes2022probing}. 
However, as well as variation in coherence times of a single qubit, the ensemble of qubits shows qubit-to-qubit variation which will have statistical distributions determined by the materials and manufacturing process used. 
The median-of-medians gives a representative number to characterise a manufacturing and assembly process. 
The highest single-shot values or best single-qubit-median~\cite{bland2025millisecond} may, and often are, presented to show what a process or platform is capable of achieving. 
The lowest coherence times are likely the cause of error-hot-spots which limit the performance of processors based on a specific platform~\cite{mohseni2024build, weeden2025statistics}. 
Given variation between qubits, it is not clear \emph{a priori} how many qubits should be measured to, with some degree of confidence, determine these metrics for a given manufacturing and assembly process.

\begin{figure}
    \centering
    \includegraphics[width=\linewidth]{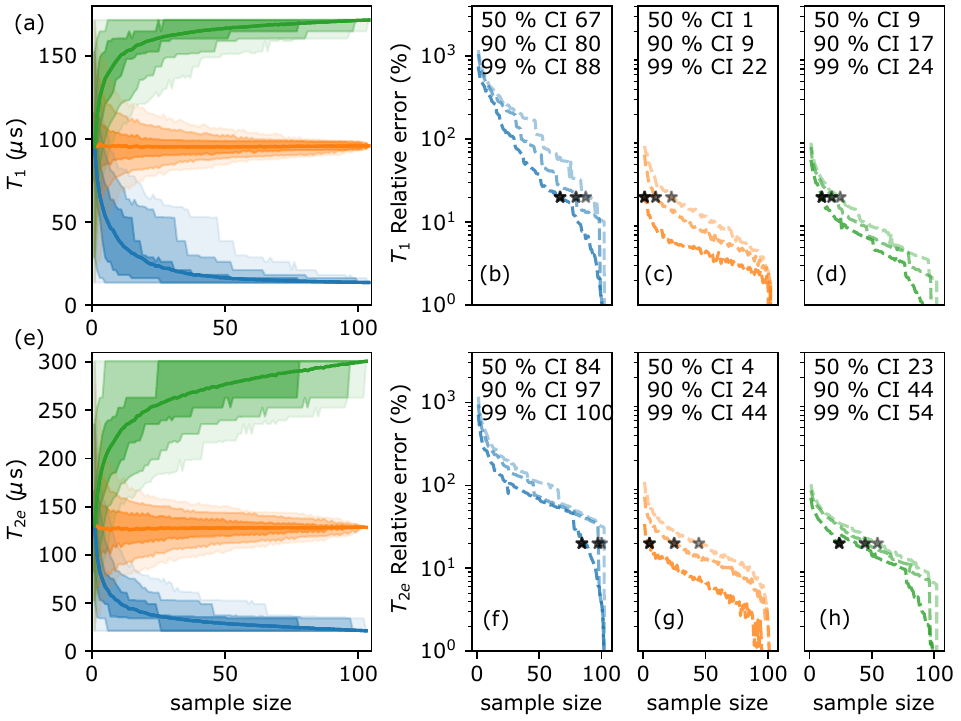}
    \caption{Bootstrapped sampling of (a) $T_1$ and (e) $T_{2e}$ from experimental measurements. The average value obtained is shown by the line, where the different shaded regions indicate a 50, 90, and 99\% confidence interval. We show values for the maximum, median and minimum value of median coherence times from different sample sizes. The same information is replotted for (b-d) $T_1$ and (f-h) $T_{2e}$ showing relative error for the different confidence intervals as a function of the sub-sample size in the bootstrapped analysis. The stars mark 20\% error for the different confidence intervals with the requisite sample sizes noted as text.}
    \label{fig:coherence_stats}
\end{figure}

In Fig.~\ref{fig:coherence_stats} we present bootstrapped sampling of the ensemble of experimentally measured median coherence times of qubits. Here we randomly select sub-samples with different sizes from the ensemble of experimental measurements of coherence times, $T_1$ and $T_{2e}$. Doing this 2000 times, we compute the median of each sub-sample and the variation, finding different quantiles for the different confidence intervals. In Fig.~\ref{fig:coherence_stats}~(a, e) we show how the mean value evolves with sample size as well as the variation which is shown by different density shading. 

The variation in results relative to the true values from our ensemble gives a relative error expected from sub-samples of different sizes shown in Fig.~\ref{fig:coherence_stats}~(b-d, f-h). We are able to resolve the median with low errors for relatively small sub-sample sizes. A sub-sample of 5 qubits gives the median value with 20\% error in a 50\% confidence interval and $<$20 qubits gives the median value with 20\% error in a 90\% confidence interval for both $T_1$ and $T_{2e}$. To resolve the minimum and maximum values with a 20\% error we require much larger sub-sample sizes, particularly the minimum value which requires a sub-sample approximately half the full sample size to measure the minimum value within 20\% in a 50\% confidence interval. This demonstrates how large N studies on the coherence of qubits are an important complement to detailed single-qubit interrogation, particularly for assessing a manufacturing process to build a quantum computer. High-throughput packages therefore offer an important function for characterising manufacturing processes. 

\section{Conclusions}
Here we present the design of a package that can house over 500 superconducting qubits. We consider multiple deleterious effects for superconducting qubits, mitigating them by the design of the package. We characterise loss rates and show that the package induces small amounts of loss, which agrees with the experimental measurements of coherence times aligned to those we would predict from dielectric loss measurements performed on devices made from a similar process. 
We demonstrate high coherence times over $>$100 qubits, with median $T_1$, $T_{2e}$ times of approximately 100$~\mu$s. Combining loss simulations and qubit coherence measurements we place a lower bound on the seam conductance of $3\times10^3~\Omega^{-1}{\rm m}^{-1}$ between aluminium and indium joints, a material interface which has previously been suggested to form an intermetallic. 
We demonstrate $T_1$-limited readout with an average readout error of 2.5\%. Well established routes to improve readout error, such as the introduction of parametric amplifiers, would further reduce this by allowing faster readout. 
By bootstrapping analysis, we show the utility of large N studies of qubit coherence, particularly identifying the best and worst performing qubits which are important numbers to understand manufacturing processes. 
The approach documented in this work is a promising route to package fault-tolerant quantum processors and is an important tool today, providing statistically rich feedback to optimize the manufacture of superconducting quantum processors. 


\begin{acknowledgments}
We thank the members of the OQC Fabrication team for the manufacture of the monolithic die used in this work. We thank the OQC Laboratory team for help with the cryostats and polishing processes performed on the packaging. 
We thank Ailsa Keyser and Apoorva Hegde for discussions on readout fidelity. 
We thank Jessica Cheung and Hilary Anderson for project management support. 
We thank Richard Bounds, Brian Vlastakis, and Peter Leek for providing critical feedback on this manuscript.
We thank the Royal Holloway University of London SuperFab Facility for their support.
\end{acknowledgments}

\bibliography{bibliography}

\newpage

\appendix

\end{document}


\preprint{APS/123-QED}

\title{Supplementary Materials for Design and Operation of Wafer-Scale Packages Containing $>$500 Superconducting Qubits}


\affiliation{%
 Oxford Quantum Circuits, Thames Valley Science Park, Shinfield, Reading, United Kingdom, RG2 9LH}%


\maketitle


\section{Spatial variation}

\begin{figure}
    \centering
    \includegraphics[width=\linewidth]{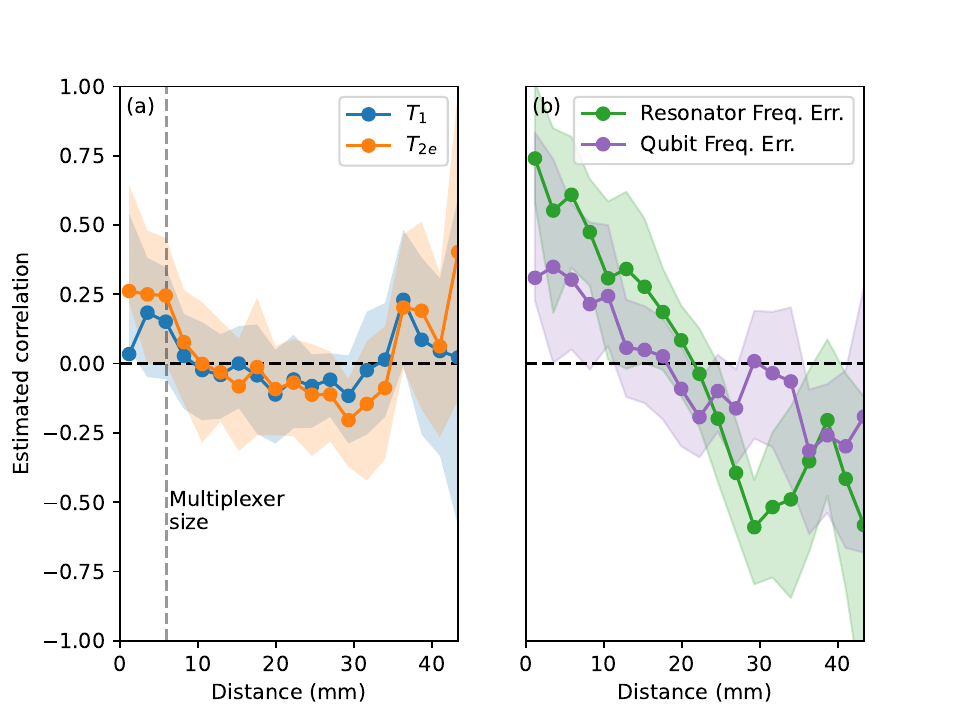}
    \caption{Pearson correlation coefficient of (a) qubit $T_1$ and $T_{2e}$ and (b) Resonator/qubit frequency errors both presented as a function of inter-qubit distance on the wafer. The shaded region indicate a 95\% confidence interval on the correlation values computed by a boot-strapping analysis. }
    \label{fig:correlation}
\end{figure}
In the main text we comment briefly on spatial correlations of coherence times - concluding that given the variance of data and the spatial distribution of qubits there are no definite trends. Understanding spatial correlations in coherence times is a powerful tool to link the figure of merit to manufacturing variations. Here we consider a Pearson correlation coefficient, testing whether the coherence times of qubits are spatially correlated. The Pearson coefficient is computed by renormalising the relevant dataset to a mean of zero and unit standard deviation before computing the correlation coefficient, i.e. the product of the renormalised values, for all pairwise combinations. The correlation values are then binned by distance. Errors are computed using a boot-strapping analysis where the same analysis is performed with sub-sets of the total dataset to establish a confidence interval.  

In Fig.~\ref{fig:correlation}~(a) we present the Pearson correlation coefficient for $T_1$ and $T_{2e}$ and contrast this to Fig.~\ref{fig:correlation}~(b) where we compute the same value for resonator and qubit frequency errors. The frequency errors are defined as the difference between a measured and a designed value. Coherence times show little correlation within errors. The very shortest distances do have a small positive correlation in $T_{2e}$ where 0 is not within a 95\% confidence interval however additional data would be needed to form strong conclusions from this analysis. We contrast this to the same analysis applied to frequency errors from qubits on the same wafer. Here we see a strong correlation at small distances. This is likely caused by variations in lithography during device manufacture where spun-resist can have thickness profiles and masked-based lithography can induce systematic variations to exposure across a wafer. We believe that these analyses will be powerful, particularly when complete datasets measuring $\sim$500 qubits are performed.

\section{RF simulations}
\begin{figure}
    \centering
    \includegraphics[width=\linewidth]{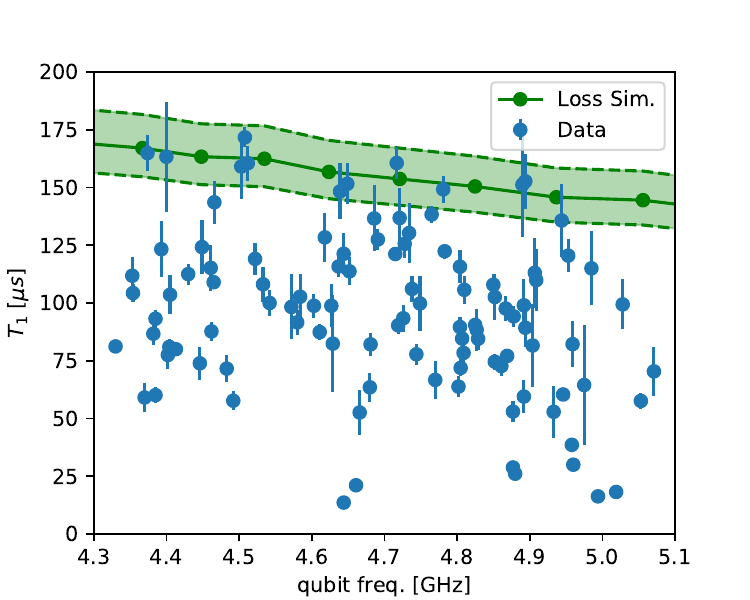}
    \caption{Frequency dependence of measured $T_1$ times. Data points are median values for the different qubits and error bars represent standard deviations of the measured $T_1$ times. Overlaid are the results of finite element simulations with the addition of loss tangents from literature with the shaded region between the dashed line indicating the quoted uncertainties.  }
    \label{fig:T1freq}
\end{figure}

We use finite element (FE) sims to infer the loss rate in our qubit geometry neglecting packaging-related losses, but instead considering losses from the superconducting device. This is shown as a range in Fig.~6~(b) of the main text and with full data and simulations in Fig.~\ref{fig:T1freq}. Here we consider loss from the material interfaces (metal air, substrate air, metal substrate) and the bulk dielectric. The simulated limit to coherence time drops with frequency. This is because of the relationship between coherence time and quality factor where $Q = \omega T_1$. While there is variation in the coherence times of qubits, none of them exceed the bounds expected from loss tangent measurements indicating that this places a bound on the coherence of the qubits. The dominant loss rate here is that arising from interfaces, with a small contribution from the low loss substrate.

\begin{figure*}
    \centering
    \includegraphics[width=0.7\linewidth]{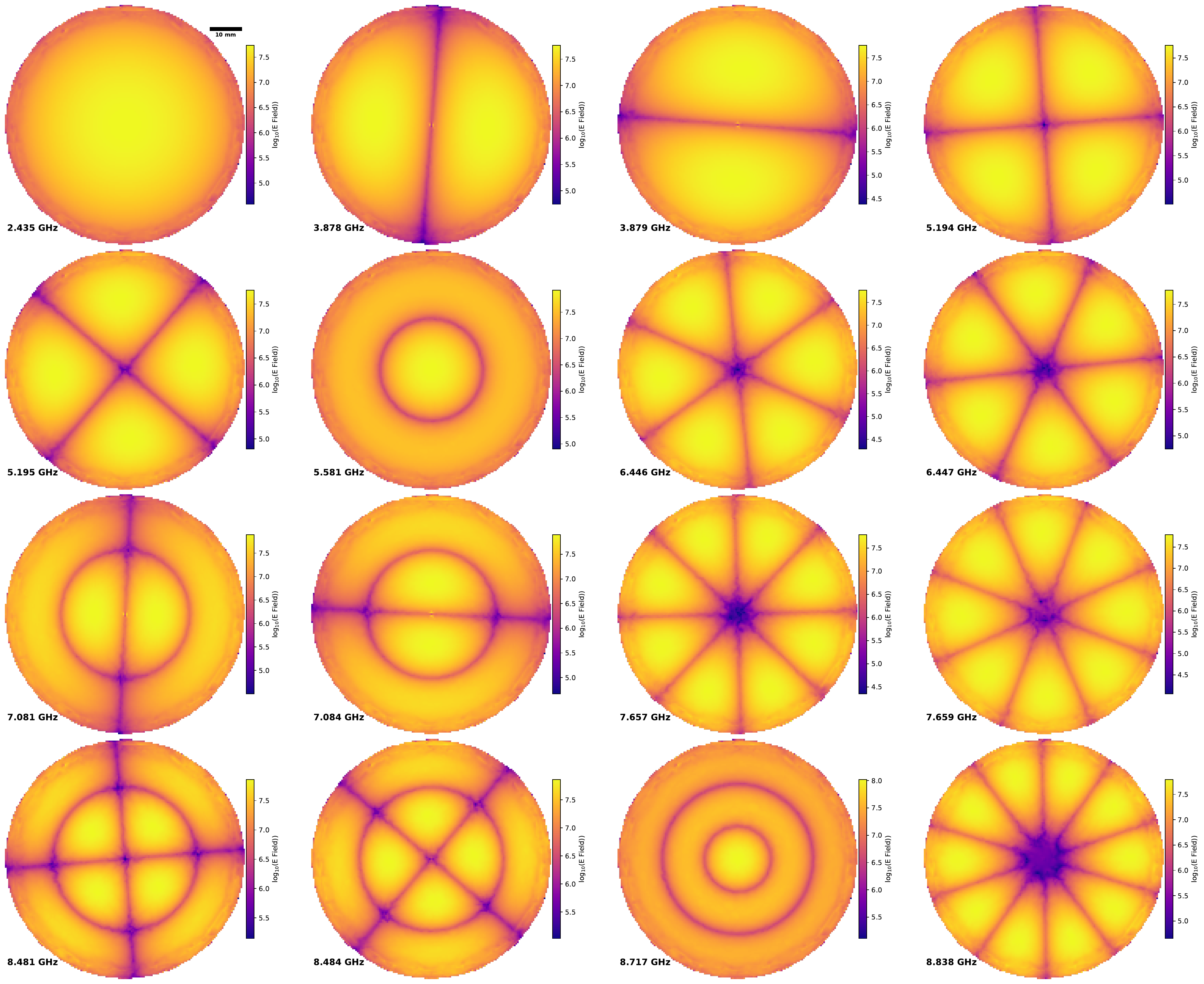}
    \caption{Electric field distributions for the first 16 Eigenmode simulations of a cavity without shunting pillars with their corresponding frequencies. }
    \label{fig:all_modes_no_pillars}
\end{figure*}

\begin{figure*}
    \centering
    \includegraphics[width=0.7\linewidth]{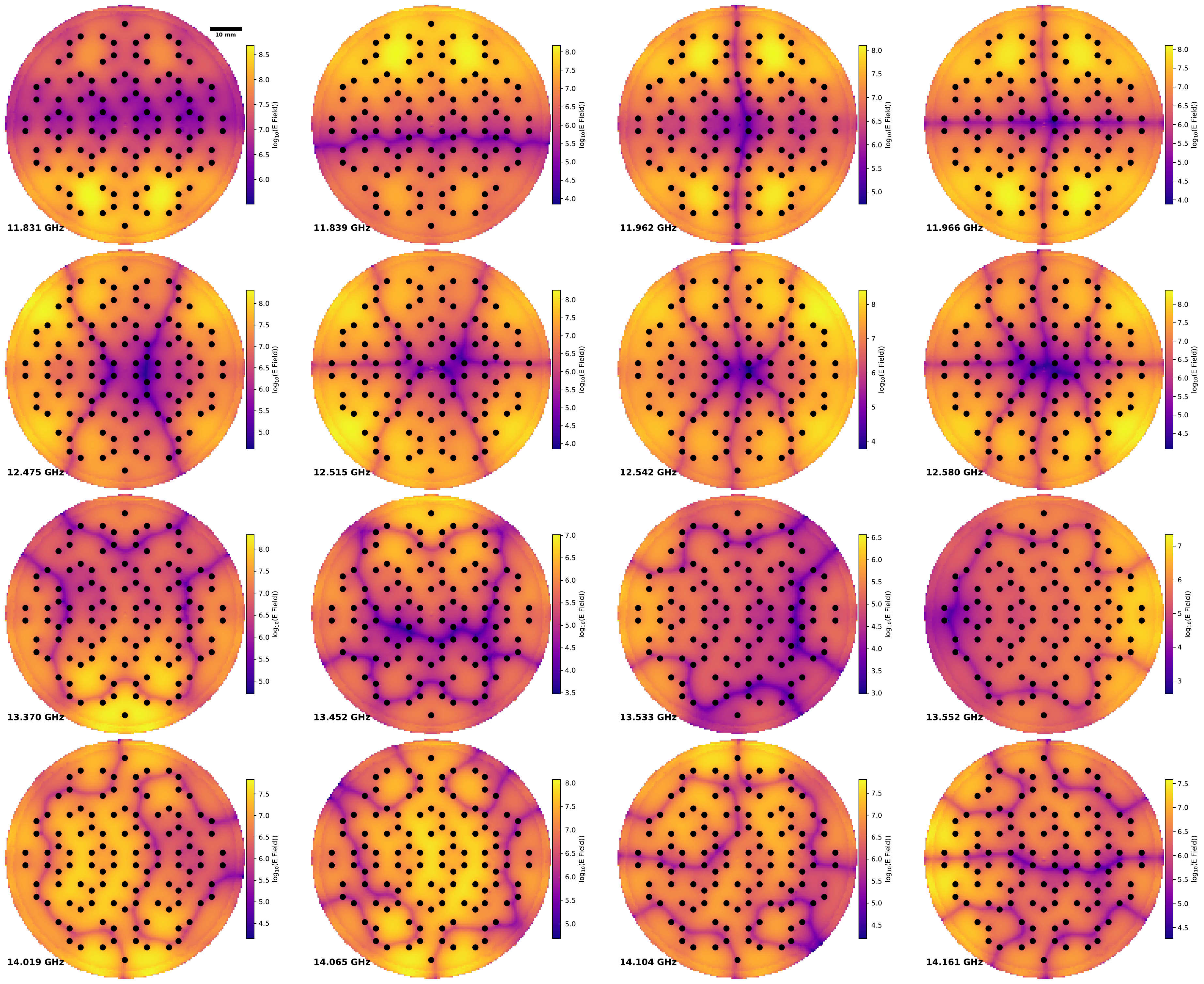}
    \caption{Electric field distributions for the first 16 Eigenmode simulations of a cavity containing shunting pillars with their corresponding frequencies. }
    \label{fig:all_modes_pillars}
\end{figure*}

As discussed in the main text, we simulate the box modes using finite element modeling. We present the E-field distribution of different microwave modes in packages without  (Fig.~\ref{fig:all_modes_no_pillars}) and with pillars (Fig.~\ref{fig:all_modes_pillars}). These simulations are performed with a dielectric wafer inside the cavity representing operational conditions. The frequencies of these (and higher modes) are shown in Fig.~2~(a) of the main text. The cavity without pillars shows a series of cylindrical cavity modes with high degrees of symmetry. Possible pillar locations are defined by the lattice of the qubits in gaps between the triangular multiplexing cells giving a regular lattice of possible pillar locations. To optimise the number of apertures that must be drilled we do not place the pillars in all possible locations, removing some pillars while ensuring that the box modes remain above frequency bands of interest. Breaking the symmetry of the pillars and the boundary conditions at the edge of the cavity mean that the modes in the cavity with shunting pillars are more complicated in their spatial variation.

The finite element simulations of the package used to compute the box modes and the loss budget require various material properties. Dielectric constants of various materials enter the finite element simulation whereas other material properties are used when computing the participation ratios from the simulations (oxide thicknesses, skin depths and penetration depths) and the arising loss rates (loss tangents, resistivity).
We tabulate the material properties used in FE sims of losses arising from qubit packaging in Table~\ref{tab:materials}. 

\begin{table}
\begin{tabular}{|l|l|l|}
\hline
\multicolumn{1}{|c|}{} & \multicolumn{1}{c|}{Value} & Ref. \\ \hline
Al Penetration depth&50~nm& \cite{faber1955penetration}\\ \hline
$R_{S,\rm Al, bulk}$&3$~\mu\Omega$ & \cite{lei2023characterization}\\ \hline
AlO$_{\rm x}$ thickness &3~nm &\cite{wenner2011surface,lei2023characterization}\\ \hline
$\epsilon_{r, \rm AlO}$ & 10 &\cite{wenner2011surface} \\ \hline
$g_{\rm seam, Al/Al}$& 700&\cite{lei2023characterization}\\ \hline
$\tan\delta_{\rm Al, bulk ,MA}$&  0.01 &\cite{lei2023characterization} \\ \hline
AgO$_{\rm x}$ thickness &  $\sim$2~nm~ & \cite{de1989oxidation} \\ \hline
$R_{S,\rm Cu, bulk}$& $\sim1~{\rm m}\Omega$ & \cite{calatroni2019cryogenic} \\ \hline
$\epsilon_{r, \rm AgO}$ &  7& \cite{petousis2017high}\\ \hline
$\epsilon_{r, \rm Rogers}$ & 2.2 & \cite{RogersData}\\ \hline
$\tan\delta_{\rm Rogers}$ & $7\times 10 ^{-4}$ & \cite{mazierska2005loss}\\ \hline
Cu skin depth & $\sim$ 840~nm & \cite{ratzinger2022anomalous}\\ \hline
\end{tabular}
\caption{Table of material properties used in loss calculations with relevant references}
\label{tab:materials}
\end{table}

When performing RF simulations of the packaging loss budget (Fig.~3 main text) we use different models to optimise computational speed. When varying the position of the qubit in the package we build a model of the full package and include one qubit/resonator pair which we place at different locations. When simulating the effects of changing the distance between the surface of the PCB and the bottom of the wafer we use a unit cell simulation. This is a smaller, cylindrical cavity which contains a single qubit/resonator pair. 

\section{Thermal Modeling}
In Table~\ref{tab:thermals} we show the heat loads on each stage of the dilution refrigerator. These are broken down by active and passive heat loads. We show the resultant temperatures of the different stages of the refrigerator determined from our model considering these different heat loads. We consider an Oxford Instruments LX Proteox 850 dilution refrigerator which has 5 stages. PT1 and PT2 are the two pulse tube plates, STL is the still plate, CLD is the cold plate and the MXC is the mixing chamber plate.

\begin{table*}[h]
\begin{tabular}{|l|lll|lll|}
\hline
      & \multicolumn{3}{c|}{High Throughput mode}                                                       & \multicolumn{3}{c|}{QPU mode}                                                         \\ \hline
Stage & \multicolumn{1}{l|}{Passive}  & \multicolumn{1}{l|}{Active/Dissipative} & Temperature & \multicolumn{1}{l|}{Passive}  & \multicolumn{1}{l|}{Active/Dissipative} & Temperature \\ \hline
PT1   & \multicolumn{1}{l|}{2.3 W}    & \multicolumn{1}{l|}{0 W}                & 35.2 K      & \multicolumn{1}{l|}{2.3 W}    & \multicolumn{1}{l|}{0 W}                & 35.1 K      \\ \hline
PT2   & \multicolumn{1}{l|}{1.4 W}    & \multicolumn{1}{l|}{1.007 W}            & 4.1 K       & \multicolumn{1}{l|}{1.4 W}    & \multicolumn{1}{l|}{993.2 mW}           & 4.1 K       \\ \hline
STL   & \multicolumn{1}{l|}{5.1 mW}   & \multicolumn{1}{l|}{0 W}                & 918.9 mK    & \multicolumn{1}{l|}{5.1 mW}   & \multicolumn{1}{l|}{0 W}                & 916.5 mK    \\ \hline
CLD   & \multicolumn{1}{l|}{1.2 mW}   & \multicolumn{1}{l|}{220.7 µW}           & 130.6 mK    & \multicolumn{1}{l|}{1.2 mW}   & \multicolumn{1}{l|}{76.4 µW}            & 127.3 mK    \\ \hline
MXC   & \multicolumn{1}{l|}{769.0 nW} & \multicolumn{1}{l|}{2.2 µW}             & 10.6 mK     & \multicolumn{1}{l|}{752.9 nW} & \multicolumn{1}{l|}{771.4 nW}           & 9.5 mK      \\ \hline
\end{tabular}
\caption{Heat loads and resultant temperatures of different stages of the dilution refrigerator from thermal modeling described in the main text. Heat loads are broken down into passive loads from mechanical interconnects between stages of the dilution refrigerator, including the wiring. Active loads are due to powered components, for instance semiconductor amplifiers. Dissipative loads are due to electrical signals passing through components which cause dissipation, such as attenuators, and convert the electrical signal into heat which is dumped at the relevant stage. We compare the high-throughput mode, where the system is set-up for feedback on device figures of merit and QPU mode where per-qubit control lines are introduced. }
\label{tab:thermals}
\end{table*}

\section{Extended Readout Data}
\begin{figure*}
    \centering
    \includegraphics[width=0.8\linewidth]{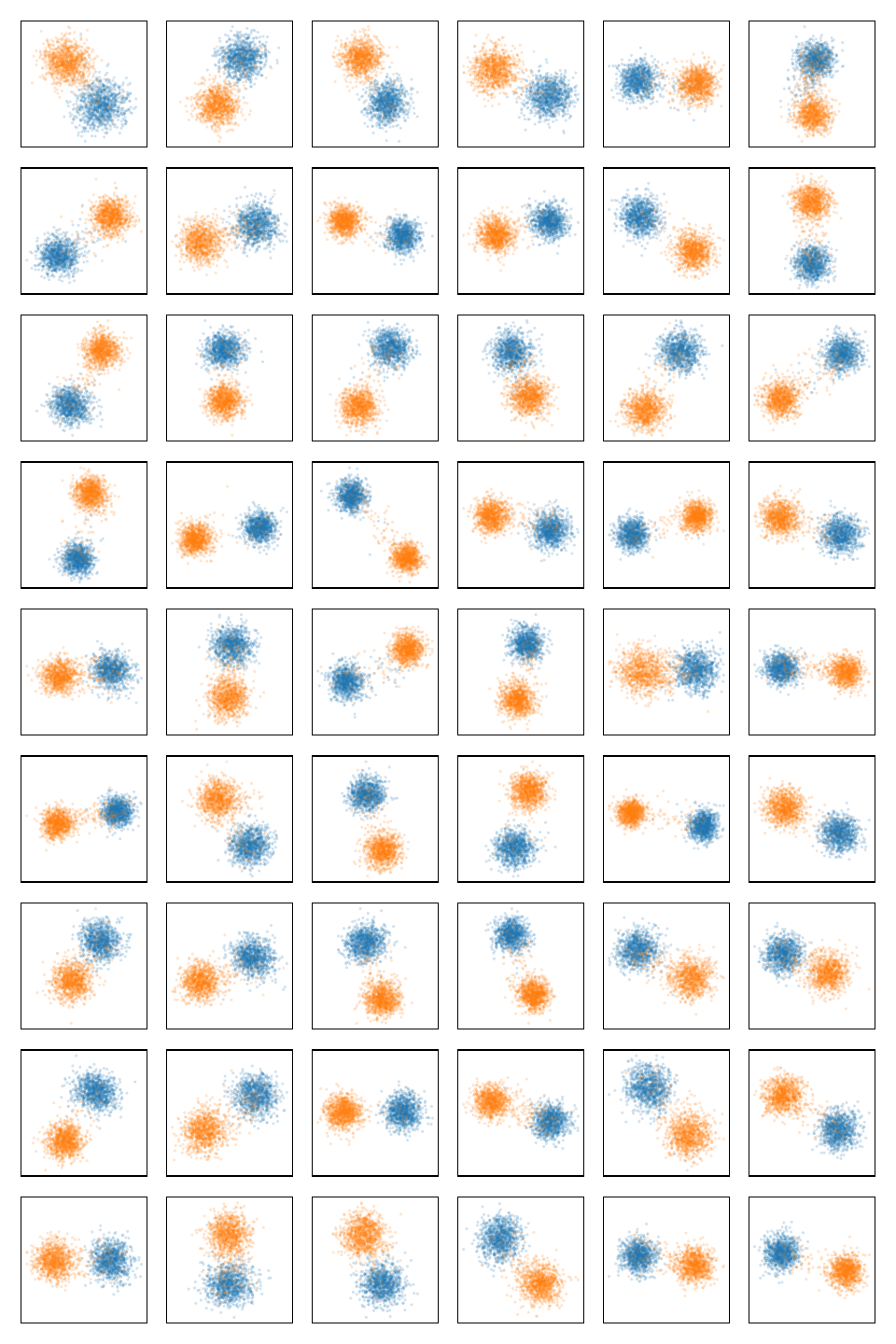}
    \caption{The raw results of optimised readout of 54 qubits with the results of readout shown on the IQ plane. Blue points indicate the qubit prepared in the ground state and orange points show the qubit prepared in the excited state. Each readout map shows well resolved clouds. }
    \label{fig:zmaps_all}
\end{figure*}

\begin{figure*}
    \centering
    \includegraphics[width=\linewidth]{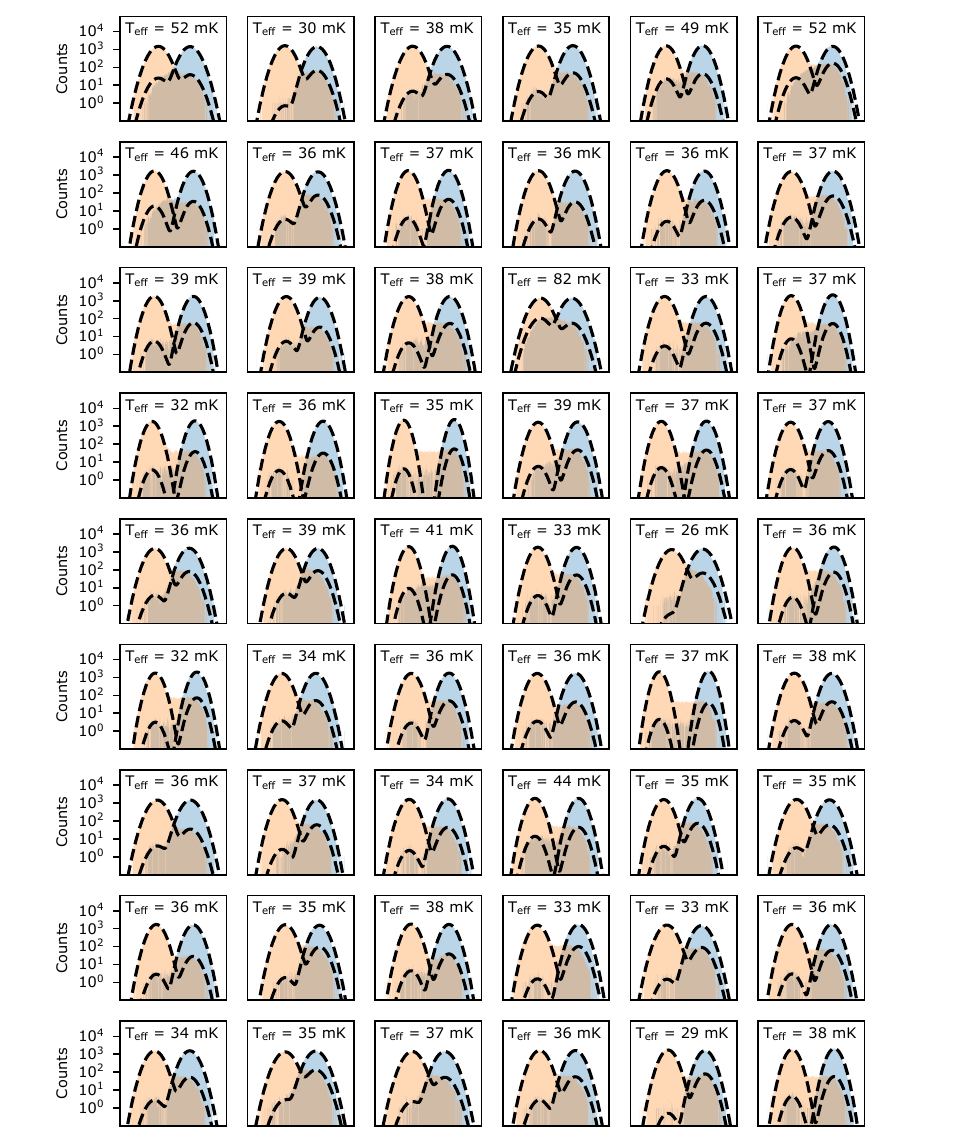}
    \caption{The results of optimised readout of 54 qubits with histograms plotted of datapoints projected onto the line of maximum state discrimination. The blue histogram is when the qubit is prepared in the ground state and the orange histogram is when it is prepared in the excited state. }
    \label{fig:zmaps_histos_all}
\end{figure*}

We show extended readout data with 54 raw measurements of readout in the IQ plane in Fig.~\ref{fig:zmaps_all} showing well separated clusters in the IQ plane. Fig.~\ref{fig:zmaps_histos_all} shows their corresponding fitted histograms. The fit to the histogram of the qubit when prepared in the ground state is used to compute effective temperature of the qubits.

\clearpage
\bibliography{bibliography}